*Research Article*

# Block Cipher's Substitution Box Generation Based on Natural Randomness in Underwater Acoustics and Knight's Tour Chain

**Muhammad Fahad Khan** ,[1,2] **Khalid Saleem** ,[1] **Tariq Shah**,[3]
**Mohammad Mazyad Hazzazi** ,[4] **Ismail Bahkali**,[5] **and Piyush Kumar Shukla**[6]

[1]*Department of Computer Science, Quaid-i-Azam University, Islamabad, Pakistan*
[2]*Department of Software Engineering, Foundation University Islamabad, Islamabad, Pakistan*
[3]*Department of Mathematics, Quaid-i-Azam University, Islamabad, Pakistan*
[4]*Department of Mathematics, College of Science, King Khalid University, Abha, Saudi Arabia*
[5]*Department of Information Sciences, King Abdulaziz University Jeddah, Jeddah 21589, Saudi Arabia*
[6]*Department of Computer Science & Engineering, University Institute of Technology,
 Rajiv Gandhi Proudyogiki Vishwavidyalaya, Bhopal, Madhya Pradesh, India*

Correspondence should be addressed to Muhammad Fahad Khan; fahad.khan@fui.edu.pk

The protection of confidential information is a global issue, and block encryption algorithms are the most reliable option for securing data. The famous information theorist, Claude Shannon, has given two desirable characteristics that should exist in a strong cipher which are substitution and permutation in their fundamental research on "Communication Theory of Secrecy Systems." block ciphers strictly follow the substitution and permutation principle in an iterative manner to generate a ciphertext. The actual strength of the block ciphers against several attacks is entirely based on its substitution characteristic, which is gained by using the substitution box (S-box). In the current literature, algebraic structure-based and chaos-based techniques are highly used for the construction of S-boxes because both these techniques have favourable features for S-box construction but also various attacks of these techniques have been identified including SAT solver, linear and differential attacks, Gröbner-based attacks, XSL attacks, interpolation attacks, XL-based attacks, finite precision effect, chaotic systems degradation, predictability, weak randomness, chaotic discontinuity, and limited control parameters. The main objective of this research is to design a novel technique for the dynamic generation of S-boxes that are safe against the cryptanalysis techniques of algebraic structure-based and chaos-based approaches. True randomness has been universally recognized as the ideal method for cipher primitives design because true random numbers are unpredictable, irreversible, and unreproducible. The biggest challenge we faced during this research was how can we generate the true random numbers and how can true random numbers utilized for strengthening the S-box construction technique. The basic concept of the proposed technique is the extraction of true random bits from underwater acoustic waves and to design a novel technique for the dynamic generation of S-boxes using the chain of knight's tour. Rather than algebraic structure- and chaos-based techniques, our proposed technique depends on inevitable high-quality randomness which exists in underwater acoustics waves. The proposed method satisfies all standard evaluation tests of S-boxes construction and true random numbers generation. Two million bits have been analyzed using the NIST randomness test suite, and the results show that underwater sound waves are an impeccable entropy source for true randomness. Additionally, our dynamically generated S-boxes have better or equal strength, over the latest published S-boxes (2020 to 2021). According to our knowledge first time, this type of research has been conducted, in which natural randomness of underwater acoustic waves has been used for the construction of block cipher's substitution box.

## 1. Introduction

Information security is the protection of secret data from illegal access, disclosure, inspection, destruction, disruption, and modification. The protection of confidential information is a global issue and block encryption algorithms are the most reliable option [1]. Block cipher is a branch of deterministic algorithm that works on the static length of bits, which is called block. Block cipher algorithms split the plaintext into various blocks of size $k$, to generate the same number of encrypted blocks of size $n$. Block ciphers encrypt one block at a time and the size of the output block is always equal to the input block and the transformation from input block to output block is done through the key whitening operation. Block cipher merged the confusion-diffusion primitives iteratively using a round function to generate an encrypted text. AES, DES, GOST, and BLOWFISH are the most prominent block ciphers of the industry that used the same strategy. For the block encryption algorithms such as AES, GOST, BLOWFISH, DES, and linear-differential attacks are the most powerful attacks [2–6]. In the differential attack, the basic purpose is to detect the sequential patterns from the encrypted text and for this purpose, the attacker tries to apply a specific set of inputs to trace the change in output. In the linear attack, the basic purpose is to find the linear relation among the plain text, cipher text with the corresponding keys. The responsibility to create a randomized relation among ciphertext and the key is on the confusion component; also, the confusion component is totally responsible to provide resistance against the linear and differential attacks [1–11]. Block cipher's confusion component is generally known as substitution (S-box) which transforms $k$ input bits into $m$ output bits through $S:\{0,1\}^k \longrightarrow \{0,1\}^m$, transforms vector $z = [z_{n-1}, z_{n-2}, z_{n-3} \ldots z0]$ into output vector $k = [k_{n-1}, k_{n-2}, k_{n-3}, \ldots, k_0]$.

As S-box is the only nonlinear primitive of block cipher, so the block cipher strength depends on its design. Cipher designers used various approaches to construct good quality S-boxes. Chaos-based and algebraic structure-based techniques are highly used for the construction of S-boxes. Chaos-based and algebraic structure-based techniques have favourable features for S-box construction, but many cryptanalysis of these techniques have been identified in the current literature. These cryptanalysis are described in Section 3. Underwater acoustic is generated by a diverse nature of sound sources such as underwater volcanoes, snapping shrimp, reverberation, vibrating objects, breaking waves, marine life, man-made sources, rain, geological activities, scattering waves, reflection waves, random motion of water molecules, lightning strikes, ice cracking, earthquake, compression, and decompression of water molecules [12–22]. Due to these diverse natures of sound sources, inevitably high-quality randomness exists in the amplitude characteristic of the underwater acoustics, which was our main source of inspiration because true randomness has been universally recognized as the ideal primitive for cryptography. True random numbers (TRNs) are unpredictable, irreversible, and unreproducible that is why cipher researchers endorsed the true random number for cryptographic primitives design [23–30].

The main idea of this research paper is extraction of true random bits from underwater acoustic waves and to design a novel technique for the dynamic generation of cryptographic S-boxes using the chain of knight's tour. The main benefit of our approach is that the proposed technique depends on the natural randomness of underwater acoustic waves for the construction of S-boxes and that's why various existing chaos and algebraic structure-based attacks are bypassed for our proposed technique.

The rest of the paper is arranged as follows. Section 2 presents our main contribution. Section 3 shows the potential cryptanalysis and attacks. Section 4 describes the proposed methodology for the dynamic generation of strong S-boxes. Section 5 presents results and discussion. Section 6 shows the conclusion.

## 2. Contribution

(i) A novel technique is proposed based on combination selection, for the generation of true random numbers from the randomness which exists in the amplitude property of underwater acoustics. As an assessment, two million bits have been analyzed using the NIST randomness test suite, and results show that underwater acoustic waves are an impeccable entropy source for TRNG.

(ii) Knight's tour-based, a novel technique is proposed, for the dynamic generation of S-boxes and as a result attacks of algebraic- and chaos-based techniques are not applicable and irrelevant for our proposed technique.

(iii) According to our knowledge first time, this type of research has been conducted, in which natural randomness of underwater acoustic waves has been used for the construction of block cipher's substitution box.

## 3. Potential Attacks of Existing S-Box Designs

As we said before, chaos-based and algebraic structures-based techniques are widely used for the construction of Shannon's confusion primitive but many attacks of these techniques have been identified in the current literature including Gröbner-based attacks [2–8], SAT solver [9–11, 31–35], linear and differential attacks [36–50], XSL attacks [51–55], interpolation attacks [51, 56–58], XL-based attacks [59–61], finite precision effect [62–67], chaotic systems degradation [61–63, 68, 69], predictability [70, 71], weak randomness [62, 63, 65, 66, 72–77], chaotic discontinuity [65–67, 72, 73], and limited control parameters [78–81].

The main objective of this research is to design a novel technique for the dynamic generation of S-boxes that are safe against the attacks of algebraic structure-based and chaos-based techniques. Rather than algebraic structure- and chaos-based techniques, our proposed technique depends on inevitable high-quality randomness which exists in underwater acoustics waves. The basic concept of the proposed technique is the extraction of true random bits from underwater acoustic waves and to design a novel technique for the dynamic generation of S-boxes using the chain of knight's tour.

## 4. Proposed Methodology

The proposed method consists of two phases, the first phase is true random numbers generation based on underwater acoustics and the second phase is dynamic generation of S-boxes based on Knight's tour chain. Architecture diagram of the proposed system is depicted in Figure 1.

*4.1. True Random Numbers Generation Based on Underwater Acoustics.* In this phase, first of all, long-term underwater acoustics recordings were acquired from the doi based dataset of the Australian Antarctic Data Centre (AADC) [82]. In the dataset, the average duration of each recording is sixty minutes. Out of thousands of long-term underwater acoustic recordings, we randomly selected the 96 long-term underwater acoustic recordings but proposed technique can take any multiple of 16 files as entropy sources. Secondly, these recordings are divided into blocks of size 16 and then, the amplitude difference of every 0.5 sec is calculated. Due to the diverse nature of sound sources, the difference of each amplitude with other amplitudes is random, and this was our main source of inspiration. Other characteristics of underwater sound like frequency and timber contain low-quality randomness that is why we chose the amplitude characteristic for this research. To calculate the amplitude differences, we used the combination selection strategy by using $n!/r!\,(n-r)!$. In our case, the value of the $n$ is 16 and the value of $r$ is 2. The entire step-by-step process of this phase from underwater acoustic files input to the random bits generation is represented in the flowchart of Figure 2. The amplitude differences calculation step is depicted in Figure 3, and here, long-term underwater acoustic recording represented as $R_1, R_2, \ldots R_{16}$. Two million bits have been analyzed using the NIST randomness test suite and shown in the Table 1, and results of the NIST tests show that underwater acoustics waves are an impeccable entropy source for true randomness. There are many random extractors based on hash functions, machine learning, chaos machine, physical unclonable functions, and probabilistic methods but among all these types of random extractors. Von Neumann random extractor is the simplest and fastest method that is why we chose Von Neumann random extractor as the post-processing method.

*4.2. Dynamic Generation of S-Boxes Based on Knight's Tour Chain.* The knight's tour is more than a 1400-year-old puzzle game whose objective is to discover the legal moves on the chessboard in the way that it explores every cell only once, and in our proposed methodology, we utilized the chain of $8 \times 8$ knight's tour for the generation of S-boxes. First of all, true random numbers are acquired and divided into blocks of 64 length size, and then, each 64 length block is converted into the $8 \times 8$ chessboard matrix. Based on the knight tour rules, we traversed each element of the chessboard matrix; however, only unique elements are considered for S-box elements, and a similar procedure is repeated for coming chessboards until the completion of required length of the S-box. Initial position of the first block of the knight's tour chain is calculated through $r =$ TRNG [0] mod 8, $c =$ TRNG [1] mod 8, and the initial positions of other knights' tour chains are dependent on the second last and the last element of the S-box, which are calculated through $r =$ S-box $[n-1]$ mod 8, $c =$ S-box $[n]$ mod 8. The entire step-by-step process of this phase from true random numbers input to dynamic S-boxes generation is represented in the flowchart of Figure 4. This phase is depicted in the following Figure 5. The reverse S-box algorithm is shown in the following (Algorithm 1). From the dynamically generated S-boxes stream, we picked two S-boxes randomly as sample which are shown in Tables 2 and 3, and their reverse S-boxes are also shown in Tables 4 and 5, respectively. The maximum nonlinearity score of our sample S-boxes is higher or equal to the recently published S-boxes (from 2020 to 2021).

## 5. Results and Evaluation

In the results and evaluation section, our proposed S-boxes are evaluated by standard S-box evaluation criteria which includes nonlinearity score, bit independence criterion, linear approximation probability, strict avalanche criterion, and differential approximation probability.

*5.1. Nonlinearity.* Among all cryptographic properties, nonlinearity is the most important one. The main purpose of S-box is to gain nonlinear change from secret message to the ciphered message. For a strong encryption scheme, the mapping between input and output in an S-box must be nonlinear. The nonlinearity of the cryptographic algorithm is represented by the nonlinearity score. Nonlinearity is defined as the smallest difference of the Boolean function to the bunch of affine functions. The nonlinearity score determine the total number of bits altered to get the closest affine function in the Boolean truth table. It calculates the distance between the set of all affine functions and Boolean function. When the initial distance is obtained, the nearest affine function is achieved by inverting the bit values in the truth table of the Boolean function. By using walsh spectrum, the nonlinearity of the Boolean function is computed through [46]:

$$N_g = 2^{n-1}\left(1 - 2^{-n} \max_{\phi \varepsilon \text{GF}(2^n)} |S(g)(\phi)|\right), \qquad (1)$$

$S_{(g)}(\varphi)$ is defined as

$$S_{(g)}(\varphi) = \sum_{\phi \in \text{GF}(2^n)} (-1)^{g(x) \oplus x.\phi}, \qquad (2)$$

where $\varphi$ is a n-bit vector and $\varphi \in \text{GF}(2^n)$. $x.\varphi$ represents the bit-wise dot product of $x$ and $\varphi$:

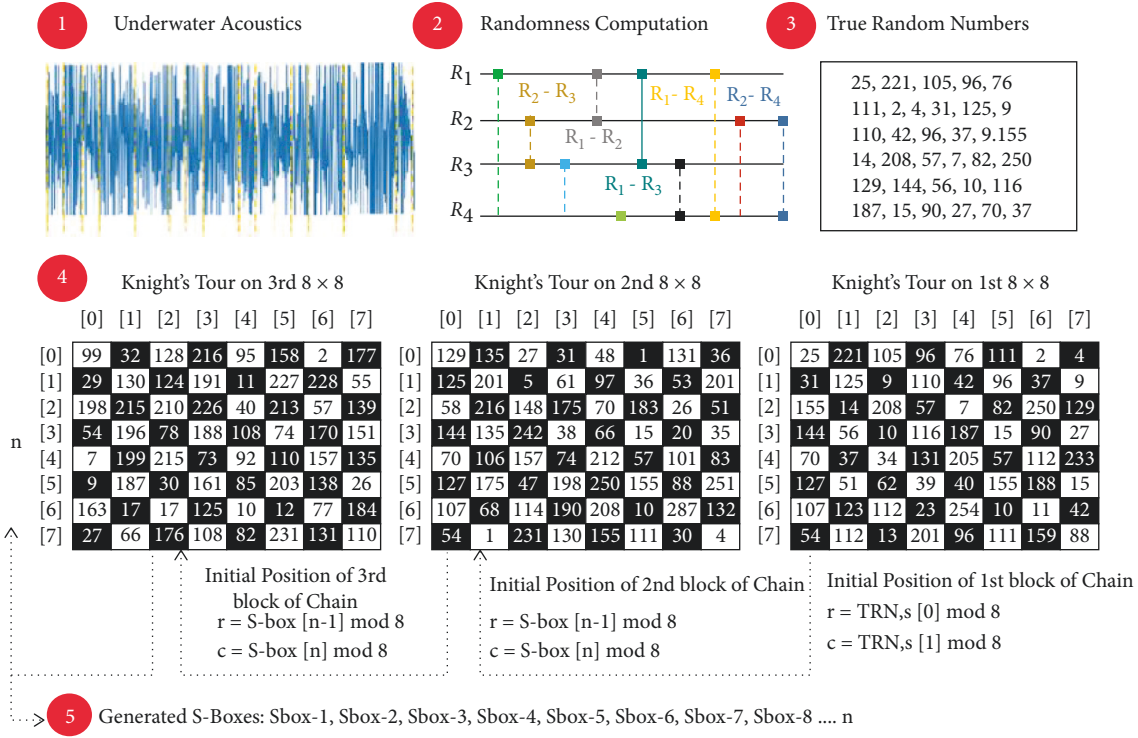

Figure 1: Architecture diagram of the proposed system.

$$x.\varphi = x1 \oplus \varphi 1 + x2 \oplus \varphi 2 \cdots + xn \oplus \varphi n. \quad (3)$$

S-box having high nonlinearity creates difficult for attacker to perform linear cryptanalysis. The maximum nonlinearity scores of our proposed Sbox-1 and Sbox-2 are 110 and 110, respectively, which is higher or equal to the recently published S-boxes. Detailed comparative analysis is shown in Table 6.

### 5.2. Strict Avalanche Criteria (SAC).

Strict avalanche criteria is the another crucial property for evaluating and according to SAC, if a single input bit is altered, all output bits will shift with probability of 1/2. SAC examined the effects of avalanche affects in encryption scheme. The modification at the input series induces a significant change in output series. SAC computes the number of output bits altered caused by inverting a single bit of input. To make the system more reliable, the output vector needed to be deviate with half probability, when one bit of input is inverted. Dependency matrix is determined to evaluate the SAC property. For an S-box that satisfies SAC property, all values were close to the ideal value of 0.5 in its dependence matrix. Dependency matrix offsets computed through equation (4) [46]. The SAC results of S-box-1 and S-box-2 are shown in Tables 7 and 8 and scores of our S-box-1 and S-box-2 are 0.495 and 0.50, respectively, which are the ideal scores for the secure S-boxes.

$$S(g) = \frac{1}{n^2} \sum_{1 \le r \le n} \sum_{1 \le w \le n} \left| \frac{1}{2} - Q_{r,w}(g) \right|, \quad (4)$$

where

$$Q_{r,w}(g) = 2^{-n} \sum_{x \in B^n} \text{gw}(x) \oplus \text{gw}(x \oplus e_r), \quad (5)$$

$e_r = [\theta r, 1\theta r, 2 \ldots \theta r, n]^T$, $[]^T$ is the transpose of matrix $\theta_{r,w} = 0, r \ne w$ or. $\theta_{r,w} = 1, r = w$.

### 5.3. BIT Independent Criterion (BIC).

BIC requires that all avalanche variables for a given set of avalanche vectors must be pair-wise independent. By modifying the input bits, BIC is used to study the behaviour of the output bits. When the output bits behave independent of one another, the S-box holds the BIC property. If any single input bit $i$ is inverted, BIC states that output bits $j$ and $k$ will alter independently. This will enhance the effectiveness of confusion function. The coefficient of correlation is used to determine the independence among pair of avalanche variables. High bit independence is required to make system design incomprehensible. The bit independence of the $j^{\text{th}}$ and $k^{\text{th}}$ bits of $B^{ei}$ is [46]: in Tables 9 and 10, we can see that our randomly picked S-box-1 and S-box-2 fully fill the BIT independent criterion.

$$\text{BIC}(b_j, b_k) = \max_{1 \le i \le n} \left| \text{corr}(b_j^{ei}, b_j^{ei}) \right|. \quad (6)$$

S-box function ($h$) is described as: $h$: $\{0, 1\}^n \longrightarrow \{0, 1\}^n$ BIC parameter for the S-box function is expressed as

$$\text{BIC}(h) = \max_{1 \le j,\ k \le n} \text{BIC}(b_j, b_k). \quad (7)$$

The change in output bits is a crucial parameter in determining the cipher's strength. When the changes in output bits contrast with the input bit sequence shows

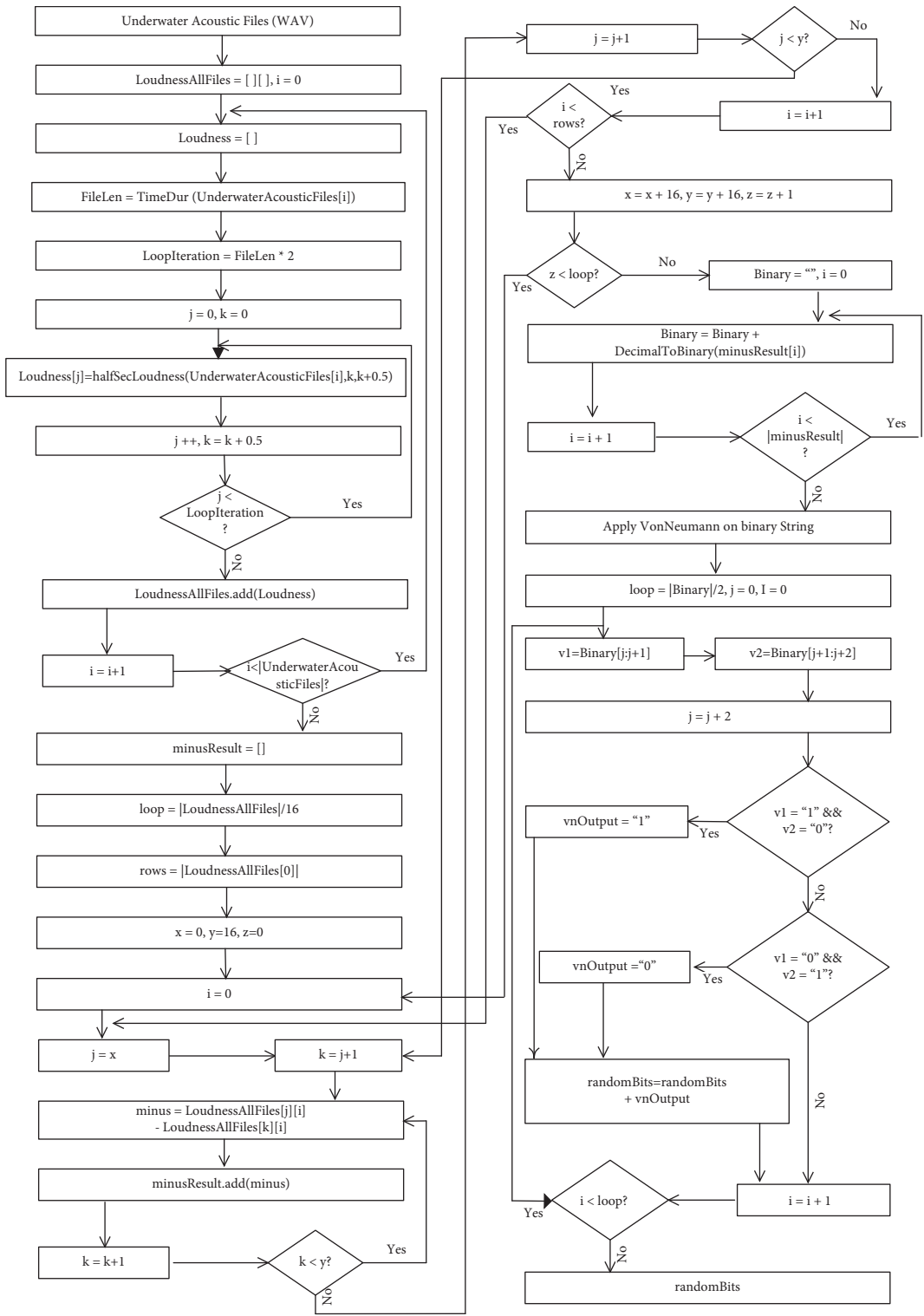

Figure 2: Flowchart of TRNG-based underwater acoustics.

sufficient independence, the mapping technique will be difficult to understand.

### 5.4. Linear Approximation Probability (LP).

LP is the cryptographic property which measures the resistance of S-box against the linear attacks. LP analysis intends to measure the maximum imbalance of the event. LP is measured by determining the total number of coincident input bits with the output bits. The input bits uniformity must be identical to the output bits. Each input bit is individually evaluated, and its results are tested in the output bits. $\gamma 1$ and $\gamma 2$ masks are selected randomly to determine the mask of all output and input values. The

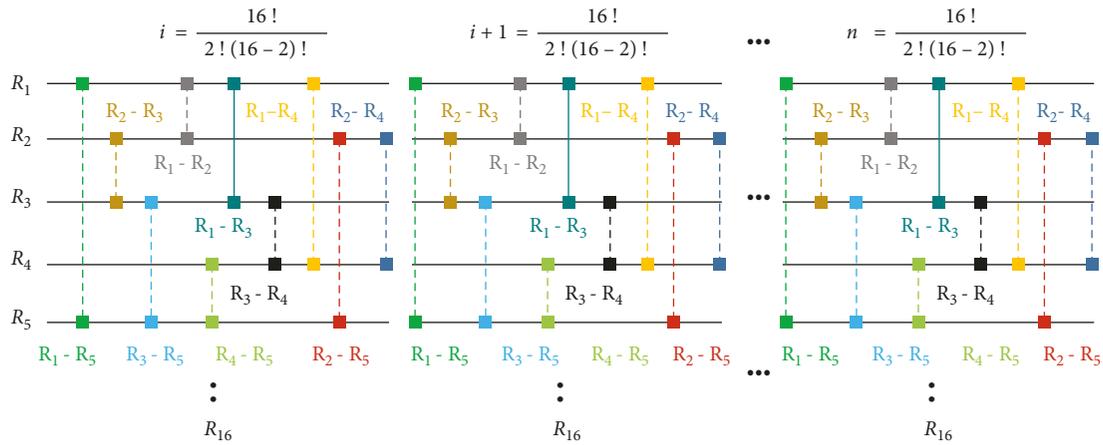

Figure 3: Computation of amplitude differences.

Table 1: Results of NIST randomness test suite.

| Type of test | | P value | Conclusion |
|---|---|---|---|
| 01. Frequency test (monobit) | | 0.8461758819031635 | Random |
| 02. Frequency test within a block | | 0.5166228701210154 | Random |
| 03. Run test | | 0.2609970420138874 | Random |
| 04. Longest run of ones in a block | | 0.34640251063536204 | Random |
| 05. Binary matrix rank test | | 0.09949346113140206 | Random |
| 06. Discrete Fourier transform (spectral) test | | 0.832838521091328 | Random |
| 07. Nonoverlapping template matching test | | 0.22184797295460632 | Random |
| 08. Overlapping template matching test | | 0.16413619193258017 | Random |
| 09. Maurer's universal statistical test | | 0.4051810932845413 | Random |
| 10. Linear complexity test | | 0.4394606534399792 | Random |
| 11. Serial test: | | 0.5703210920746249 | Random |
| | | 0.5412977586951687 | Random |
| 12. Approximate entropy test | | 0.013704869478928823 | Random |
| 13. Cummulative sums (forward) test | | 0.9081561792752144 | Random |
| 14. Cummulative sums (reverse) test | | 0.7420961383854099 | Random |
| | State | Chi squared | P value | Conclusion |
| | −4 | 3.1940558536339703 | 0.6700965356355721 | Random |
| | −3 | 8.322704313725493 | 0.13932469392722086 | Random |
| | −2 | 2.215379649802308 | 0.8186113923928053 | Random |
| 15. Random excursions test: | −1 | 10.373856209150327 | 0.06530927491189864 | Random |
| | +1 | 5.49281045751634 | 0.3587348843928551 | Random |
| | +2 | 1.978633099330267 | 0.8520935894148005 | Random |
| | +3 | 0.872903529411762 | 0.9721522155809542 | Random |
| | +4 | 2.2030722493078865 | 0.8203920781040164 | Random |
| | State | Counts | P value | Conclusion |
| | −9.0 | 1743 | 0.3503620748973999 | Random |
| | −8.0 | 1791 | 0.22313138786599762 | Random |
| | −7.0 | 1861 | 0.09700096546916874 | Random |
| | −6.0 | 1837 | 0.09426256021374013 | Random |
| | −5.0 | 1755 | 0.17515792247265105 | Random |
| | −4.0 | 1675 | 0.3218141454622986 | Random |
| | −3.0 | 1588 | 0.6391395377015844 | Random |
| | −2.0 | 1605 | 0.43375610043914314 | Random |
| 16. Random excursions variant test: | −1.0 | 1572 | 0.4476990724652935 | Random |
| | +1.0 | 1601 | 0.19931513588782468 | Random |
| | +2.0 | 1629 | 0.30147752489003166 | Random |
| | +3.0 | 1609 | 0.523032983174088 | Random |
| | +4.0 | 1609 | 0.589348273539888 | Random |
| | +5.0 | 1662 | 0.426374068680618 | Random |
| | +6.0 | 1783 | 0.1678955379041649 | Random |
| | +7.0 | 1876 | 0.0827802734496795 | Random |
| | +8.0 | 1869 | 0.1135773370125223 | Random |
| | +9.0 | 1818 | 0.20668955769990105 | Random |

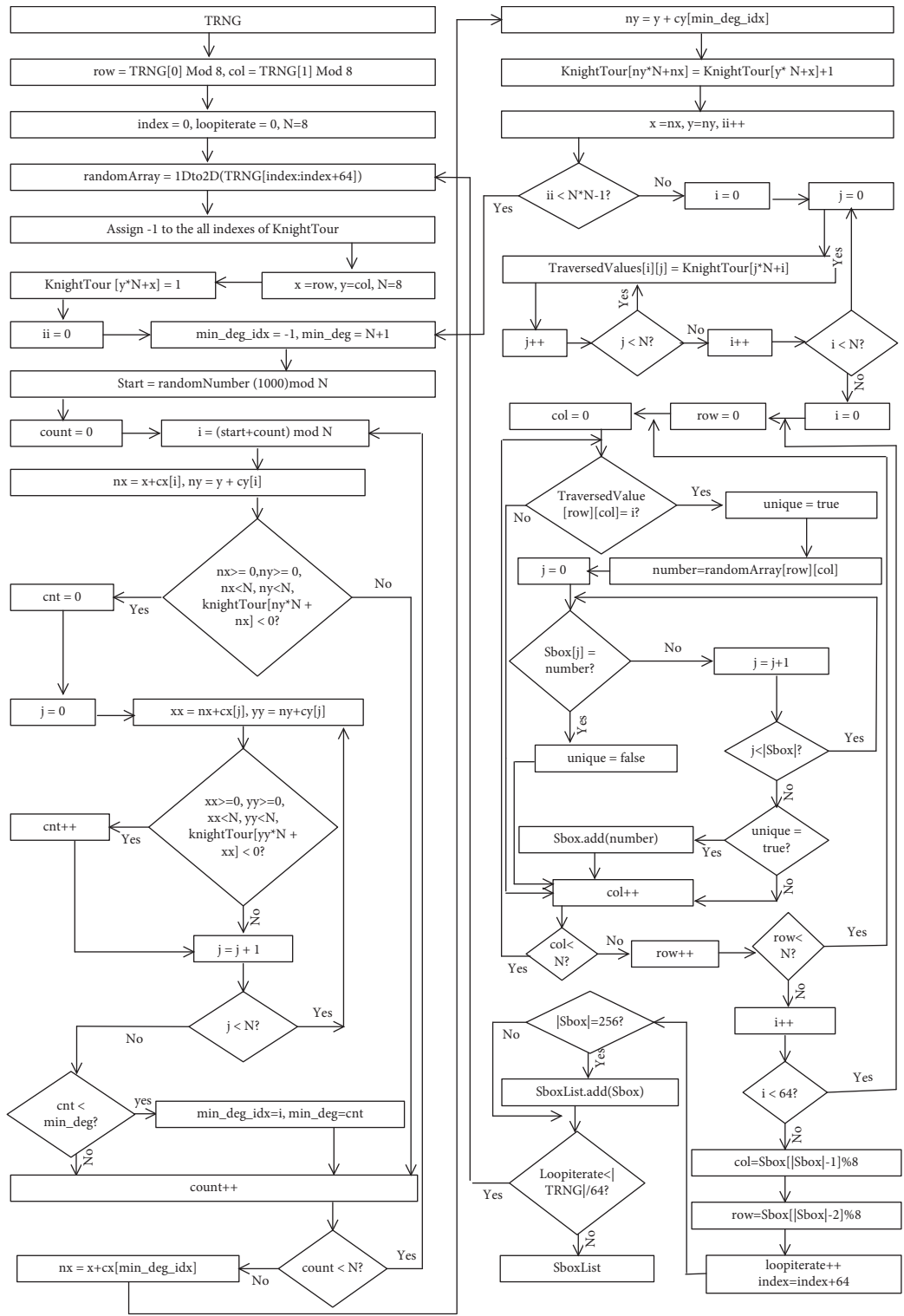

Figure 4: Flowchart of dynamic S-boxes generation based on knight's tour chain.

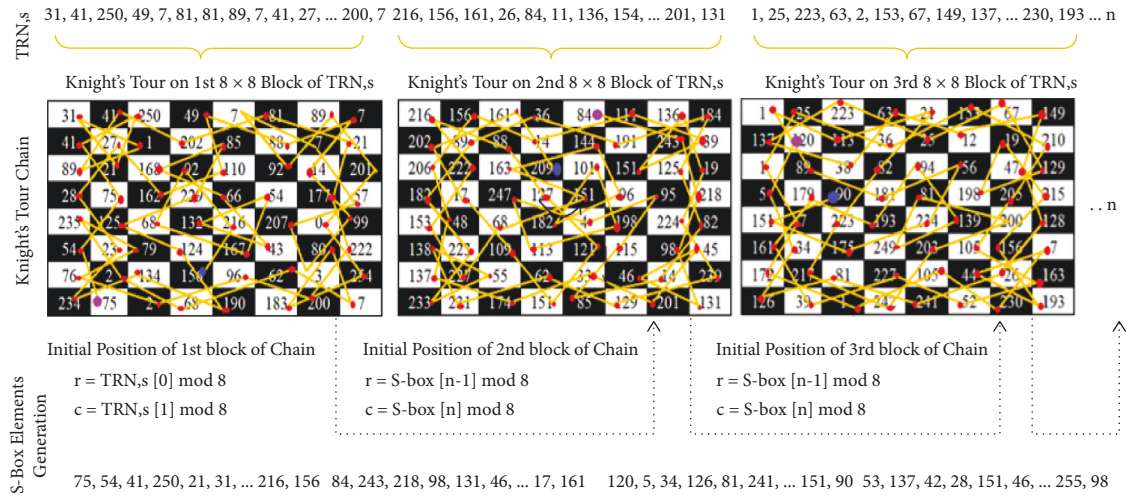

Figure 5: Dynamic generation of S-boxes based on knight's tour chain.

Table 2: Substitution box 1.

| | | | | | | | | | | | | | | | |
|---|---|---|---|---|---|---|---|---|---|---|---|---|---|---|---|
| 106 | 220 | 5 | 24 | 1 | 124 | 20 | 104 | 96 | 88 | 240 | 170 | 9 | 115 | 246 | 86 |
| 197 | 230 | 174 | 155 | 76 | 185 | 175 | 31 | 142 | 103 | 239 | 122 | 40 | 113 | 208 | 228 |
| 78 | 21 | 218 | 29 | 110 | 85 | 43 | 70 | 27 | 120 | 66 | 28 | 189 | 126 | 36 | 232 |
| 138 | 165 | 234 | 16 | 243 | 23 | 160 | 235 | 97 | 48 | 90 | 101 | 98 | 250 | 6 | 45 |
| 38 | 73 | 141 | 53 | 81 | 71 | 203 | 206 | 2 | 135 | 252 | 111 | 145 | 92 | 238 | 63 |
| 130 | 186 | 180 | 123 | 192 | 4 | 251 | 89 | 196 | 84 | 58 | 143 | 32 | 59 | 82 | 198 |
| 112 | 224 | 247 | 64 | 177 | 178 | 148 | 184 | 233 | 200 | 222 | 107 | 105 | 195 | 201 | 187 |
| 154 | 236 | 163 | 109 | 219 | 254 | 137 | 210 | 241 | 204 | 212 | 139 | 34 | 248 | 249 | 74 |
| 202 | 253 | 52 | 47 | 226 | 19 | 12 | 3 | 114 | 207 | 118 | 171 | 91 | 193 | 217 | 144 |
| 169 | 237 | 13 | 57 | 131 | 30 | 121 | 95 | 33 | 14 | 199 | 119 | 146 | 100 | 166 | 182 |
| 255 | 72 | 215 | 209 | 188 | 77 | 99 | 35 | 116 | 242 | 18 | 87 | 132 | 102 | 158 | 152 |
| 150 | 62 | 211 | 55 | 164 | 80 | 162 | 125 | 225 | 133 | 183 | 117 | 179 | 51 | 205 | 60 |
| 65 | 8 | 15 | 213 | 69 | 223 | 41 | 54 | 176 | 46 | 244 | 194 | 0 | 156 | 172 | 39 |
| 56 | 161 | 227 | 147 | 93 | 129 | 67 | 168 | 221 | 245 | 25 | 136 | 17 | 214 | 128 | 167 |
| 61 | 229 | 22 | 11 | 153 | 94 | 149 | 151 | 50 | 216 | 49 | 159 | 37 | 10 | 127 | 7 |
| 26 | 83 | 44 | 134 | 42 | 231 | 79 | 75 | 68 | 108 | 173 | 157 | 140 | 191 | 181 | 190 |

Table 3: Substitution box 2.

| | | | | | | | | | | | | | | | |
|---|---|---|---|---|---|---|---|---|---|---|---|---|---|---|---|
| 254 | 240 | 187 | 11 | 151 | 155 | 153 | 100 | 103 | 201 | 144 | 0 | 72 | 14 | 158 | 63 |
| 180 | 209 | 138 | 2 | 169 | 27 | 60 | 186 | 21 | 97 | 52 | 109 | 251 | 248 | 95 | 19 |
| 124 | 71 | 10 | 107 | 58 | 210 | 26 | 203 | 90 | 168 | 121 | 250 | 66 | 226 | 50 | 104 |
| 176 | 46 | 65 | 93 | 6 | 183 | 245 | 134 | 86 | 216 | 7 | 44 | 238 | 207 | 16 | 110 |
| 202 | 99 | 17 | 165 | 217 | 167 | 80 | 55 | 128 | 82 | 75 | 200 | 40 | 182 | 147 | 174 |
| 196 | 156 | 120 | 192 | 116 | 136 | 164 | 188 | 48 | 5 | 152 | 166 | 33 | 62 | 230 | 137 |
| 12 | 9 | 102 | 61 | 223 | 54 | 159 | 34 | 59 | 246 | 195 | 213 | 170 | 51 | 253 | 229 |
| 126 | 122 | 140 | 241 | 98 | 77 | 237 | 179 | 47 | 191 | 30 | 130 | 118 | 185 | 224 | 243 |
| 45 | 36 | 227 | 149 | 106 | 239 | 68 | 221 | 189 | 219 | 150 | 108 | 13 | 161 | 154 | 112 |
| 242 | 172 | 23 | 178 | 135 | 131 | 160 | 231 | 129 | 244 | 31 | 255 | 173 | 39 | 233 | 205 |
| 198 | 89 | 20 | 18 | 215 | 8 | 249 | 139 | 181 | 212 | 53 | 163 | 157 | 127 | 208 | 64 |
| 105 | 85 | 142 | 184 | 145 | 29 | 37 | 175 | 111 | 125 | 222 | 4 | 117 | 232 | 76 | 87 |
| 84 | 35 | 42 | 123 | 49 | 235 | 24 | 92 | 101 | 91 | 204 | 194 | 79 | 133 | 96 | 32 |
| 15 | 69 | 67 | 146 | 190 | 88 | 83 | 1 | 141 | 119 | 177 | 234 | 132 | 38 | 74 | 236 |
| 252 | 3 | 28 | 78 | 22 | 220 | 57 | 43 | 211 | 225 | 199 | 41 | 94 | 197 | 143 | 70 |
| 206 | 73 | 162 | 56 | 25 | 228 | 218 | 81 | 115 | 114 | 171 | 113 | 148 | 193 | 247 | 214 |

Table 4: Reverse S-box 1.

| | | | | | | | | | | | | | | | |
|---|---|---|---|---|---|---|---|---|---|---|---|---|---|---|---|
| 204 | 4 | 72 | 135 | 85 | 2 | 62 | 239 | 193 | 12 | 237 | 227 | 134 | 146 | 153 | 194 |
| 51 | 220 | 170 | 133 | 6 | 33 | 226 | 53 | 3 | 218 | 240 | 40 | 43 | 35 | 149 | 23 |
| 92 | 152 | 124 | 167 | 46 | 236 | 64 | 207 | 28 | 198 | 244 | 38 | 242 | 63 | 201 | 131 |
| 57 | 234 | 232 | 189 | 130 | 67 | 199 | 179 | 208 | 147 | 90 | 93 | 191 | 224 | 177 | 79 |
| 99 | 192 | 42 | 214 | 248 | 196 | 39 | 69 | 161 | 65 | 127 | 247 | 20 | 165 | 32 | 246 |
| 181 | 68 | 94 | 241 | 89 | 37 | 15 | 171 | 9 | 87 | 58 | 140 | 77 | 212 | 229 | 151 |
| 8 | 56 | 60 | 166 | 157 | 59 | 173 | 25 | 7 | 108 | 0 | 107 | 249 | 115 | 36 | 75 |
| 96 | 29 | 136 | 13 | 168 | 187 | 138 | 155 | 41 | 150 | 27 | 83 | 5 | 183 | 45 | 238 |
| 222 | 213 | 80 | 148 | 172 | 185 | 243 | 73 | 219 | 118 | 48 | 123 | 252 | 66 | 24 | 91 |
| 143 | 76 | 156 | 211 | 102 | 230 | 176 | 231 | 175 | 228 | 112 | 19 | 205 | 251 | 174 | 235 |
| 54 | 209 | 182 | 114 | 180 | 49 | 158 | 223 | 215 | 144 | 11 | 139 | 206 | 250 | 18 | 22 |
| 200 | 100 | 101 | 188 | 82 | 254 | 159 | 186 | 103 | 21 | 81 | 111 | 164 | 44 | 255 | 253 |
| 84 | 141 | 203 | 109 | 88 | 16 | 95 | 154 | 105 | 110 | 128 | 70 | 121 | 190 | 71 | 137 |
| 30 | 163 | 119 | 178 | 122 | 195 | 221 | 162 | 233 | 142 | 34 | 116 | 1 | 216 | 106 | 197 |
| 97 | 184 | 132 | 210 | 31 | 225 | 17 | 245 | 47 | 104 | 50 | 55 | 113 | 145 | 78 | 26 |
| 10 | 120 | 169 | 52 | 202 | 217 | 14 | 98 | 125 | 126 | 61 | 86 | 74 | 129 | 117 | 160 |

Table 5: Reverse S-box 2.

| | | | | | | | | | | | | | | | |
|---|---|---|---|---|---|---|---|---|---|---|---|---|---|---|---|
| 11 | 215 | 19 | 225 | 187 | 89 | 52 | 58 | 165 | 97 | 34 | 3 | 96 | 140 | 13 | 208 |
| 62 | 66 | 163 | 31 | 162 | 24 | 228 | 146 | 198 | 244 | 38 | 21 | 226 | 181 | 122 | 154 |
| 207 | 92 | 103 | 193 | 129 | 182 | 221 | 157 | 76 | 235 | 194 | 231 | 59 | 128 | 49 | 120 |
| 88 | 196 | 46 | 109 | 26 | 170 | 101 | 71 | 243 | 230 | 36 | 104 | 22 | 99 | 93 | 15 |
| 175 | 50 | 44 | 210 | 134 | 209 | 239 | 33 | 12 | 241 | 222 | 74 | 190 | 117 | 227 | 204 |
| 70 | 247 | 73 | 214 | 192 | 177 | 56 | 191 | 213 | 161 | 40 | 201 | 199 | 51 | 236 | 30 |
| 206 | 25 | 116 | 65 | 7 | 200 | 98 | 8 | 47 | 176 | 132 | 35 | 139 | 27 | 63 | 184 |
| 143 | 251 | 249 | 248 | 84 | 188 | 124 | 217 | 82 | 42 | 113 | 195 | 32 | 185 | 112 | 173 |
| 72 | 152 | 123 | 149 | 220 | 205 | 55 | 148 | 85 | 95 | 18 | 167 | 114 | 216 | 178 | 238 |
| 10 | 180 | 211 | 78 | 252 | 131 | 138 | 4 | 90 | 6 | 142 | 5 | 81 | 172 | 14 | 102 |
| 150 | 141 | 242 | 171 | 86 | 67 | 91 | 69 | 41 | 20 | 108 | 250 | 145 | 156 | 79 | 183 |
| 48 | 218 | 147 | 119 | 16 | 168 | 77 | 53 | 179 | 125 | 23 | 2 | 87 | 136 | 212 | 121 |
| 83 | 253 | 203 | 106 | 80 | 237 | 160 | 234 | 75 | 9 | 64 | 39 | 202 | 159 | 240 | 61 |
| 174 | 17 | 37 | 232 | 169 | 107 | 255 | 164 | 57 | 68 | 246 | 137 | 229 | 135 | 186 | 100 |
| 126 | 233 | 45 | 130 | 245 | 111 | 94 | 151 | 189 | 158 | 219 | 197 | 223 | 118 | 60 | 133 |
| 1 | 115 | 144 | 127 | 153 | 54 | 105 | 254 | 29 | 166 | 43 | 28 | 224 | 110 | 0 | 155 |

```
in: 2D array of integers, sbox [16, 16];
out: 2D array of integers, ReverseSbox [16, 16];
(1) ReverseSbox ⟶ |16||16|
(2) for row ⟶ 0 . . . (16) do
(3)    for col ⟶ 0 . . . (16) do
(4)       rowIS ⟶ sbox row, col div 16
(5)       colIS ⟶ sbox row, col mod 16
(6)       value ⟶ row ∗ 16 + col
(7)       ReverseSbox rowIS, colIS ⟶value
(8)    end for
(9) end for
(10) return ReverseSbox
```

Algorithm 1: ReverseSbox (S-box).

Table 6: Nonlinearity of state-of-the-art S-boxes.

| Recently published S-boxes | Maximum nonlinearity achieved |
|---|---|
| [83], 2021 | −108 |
| [85], 2021 | −110 |
| [87], 2021 | −108 |
| [89], 2020 | −108 |
| [91], 2020 | −110 |
| [93], 2020 | −108 |
| [95], 2020 | −108 |
| [97], 2020 | −108 |
| [99], 2020 | −104 |
| [99], 2020 | −108 |
| [98], 2020 | −106 |
| [101], 2020 | −108 |
| [102], 2020 | −110 |
| [104], 2021 | −108 |
| [105], 2021 | −110 |
| [84], 2021 | −110 |
| [86], 2021 | −110 |
| [88], 2021 | −108 |
| [90], 2021 | −110 |
| [92], 2020 | −102 |
| [94], 2020 | −107 |
| [96], 2020 | −104 |
| [98], 2020 | −106 |
| [100], 2020 | −105 |
| [101], 2020 | −106 |
| [95], 2020 | −108 |
| [93], 2020 | −108 |
| [103], 2020 | −108 |
| [104], 2021 | −108 |
| [106], 2021 | −108 |

Table 7: SAC results of S-box-1.

| 0.500000 | 0.562500 | 0.468750 | 0.453125 | 0.500000 | 0.421875 | 0.453125 | 0.500000 |
|---|---|---|---|---|---|---|---|
| 0.437500 | 0.515625 | 0.468750 | 0.468750 | 0.515625 | 0.500000 | 0.546875 | 0.437500 |
| 0.468750 | 0.546875 | 0.484375 | 0.515625 | 0.500000 | 0.531250 | 0.546875 | 0.500000 |
| 0.453125 | 0.500000 | 0.500000 | 0.500000 | 0.484375 | 0.453125 | 0.515625 | 0.546875 |
| 0.468750 | 0.562500 | 0.500000 | 0.500000 | 0.484375 | 0.437500 | 0.484375 | 0.500000 |
| 0.406250 | 0.546875 | 0.593750 | 0.484375 | 0.453125 | 0.390625 | 0.531250 | 0.500000 |
| 0.437500 | 0.484375 | 0.578125 | 0.453125 | 0.515625 | 0.546875 | 0.437500 | 0.484375 |
| 0.546875 | 0.515625 | 0.531250 | 0.500000 | 0.562500 | 0.437500 | 0.515625 | 0.515625 |

Table 8: SAC results of S-box-2.

| 0.531250 | 0.546875 | 0.546875 | 0.468750 | 0.421875 | 0.437500 | 0.546875 | 0.500000 |
|---|---|---|---|---|---|---|---|
| 0.546875 | 0.531250 | 0.406250 | 0.484375 | 0.562500 | 0.468750 | 0.484375 | 0.453125 |
| 0.515625 | 0.484375 | 0.500000 | 0.578125 | 0.640625 | 0.515625 | 0.546875 | 0.437500 |
| 0.562500 | 0.468750 | 0.453125 | 0.437500 | 0.500000 | 0.546875 | 0.546875 | 0.546875 |
| 0.593750 | 0.546875 | 0.531250 | 0.593750 | 0.500000 | 0.500000 | 0.468750 | 0.531250 |
| 0.500000 | 0.468750 | 0.531250 | 0.531250 | 0.437500 | 0.484375 | 0.484375 | 0.484375 |
| 0.484375 | 0.421875 | 0.546875 | 0.484375 | 0.437500 | 0.515625 | 0.515625 | 0.546875 |
| 0.500000 | 0.453125 | 0.578125 | 0.468750 | 0.562500 | 0.531250 | 0.562500 | 0.421875 |

Table 9: BIC independent matrix of S-box-1.

| — | 0.480469 | 0.484375 | 0.464844 | 0.509766 | 0.507812 | 0.517578 | 0.521484 |
|---|---|---|---|---|---|---|---|
| 0.480469 | — | 0.511719 | 0.513672 | 0.484375 | 0.486328 | 0.476562 | 0.494141 |
| 0.484375 | 0.511719 | — | 0.498047 | 0.507812 | 0.494141 | 0.503906 | 0.486328 |
| 0.464844 | 0.513672 | 0.498047 | — | 0.494141 | 0.505859 | 0.501953 | 0.496094 |
| 0.509766 | 0.484375 | 0.507812 | 0.494141 | — | 0.509766 | 0.480469 | 0.470703 |
| 0.507812 | 0.486328 | 0.494141 | 0.505859 | 0.509766 | — | 0.494141 | 0.498047 |
| 0.517578 | 0.476562 | 0.503906 | 0.501953 | 0.480469 | 0.494141 | — | 0.509766 |
| 0.521484 | 0.494141 | 0.486328 | 0.496094 | 0.470703 | 0.498047 | 0.509766 | — |

Table 10: BIC independent matrix of S-box-2.

| — | 0.501953 | 0.498047 | 0.501953 | 0.488281 | 0.529297 | 0.486328 | 0.484375 |
|---|---|---|---|---|---|---|---|
| 0.501953 | — | 0.500000 | 0.501953 | 0.484375 | 0.513672 | 0.466797 | 0.509766 |
| 0.498047 | 0.500000 | — | 0.507812 | 0.527344 | 0.474609 | 0.507812 | 0.486328 |
| 0.501953 | 0.501953 | 0.507812 | — | 0.519531 | 0.521484 | 0.494141 | 0.511719 |
| 0.488281 | 0.484375 | 0.527344 | 0.519531 | — | 0.523438 | 0.515625 | 0.521484 |
| 0.529297 | 0.513672 | 0.474609 | 0.521484 | 0.523438 | — | 0.478516 | 0.503906 |
| 0.486328 | 0.466797 | 0.507812 | 0.494141 | 0.515625 | 0.478516 | — | 0.519531 |
| 0.484375 | 0.509766 | 0.486328 | 0.511719 | 0.521484 | 0.503906 | 0.519531 | — |

mathematical expression of determining linear approximation probability is as follows [46]: the maximum LP of S-box-1 and S-box-2 is 0.125, which is also satisfies LP criteria.

$$\text{LP}_f = \max_{\gamma 1, \gamma 2 \neq 0} \left| \frac{\{x \varepsilon X | x.\gamma 1 = S(x).\gamma 2\}}{2^n} - \frac{1}{2} \right|, \quad (8)$$

where $\gamma 1$ and $\gamma 2$ represents the input and output mask in the above expression. Linear approximation probability is calculated by using these masks. $X$ represents the set of all possible inputs and $2^n$ is the total number of elements in the set. S-box with low LP value is robust enough against different linear approximation attacks.

5.5. Differential Approximation Probability (DP). The resistance of S-box to the differential attacks is assessed through the DP. DP is the probability of particular change in output bits caused by the change in input bits. An S-box must possess differential uniformity which means that each input differential is connected to the specific output differential. The XOR values of all output must have equal probability to the XOR values of all input. The differential uniformity is measured by given expression [46]:

$$\text{DP}(\Delta x \longrightarrow \Delta y) = \left[ \frac{\#\{x \varepsilon X | (S(x) \oplus S(x \oplus \Delta x) = \Delta y\}}{2^n} \right], \quad (9)$$

where $X$ represents the set of all possible input values and $2^n$ is the total number of elements in set. The maximum differential probability value a system could achieve is 4/256. The lowest value of DP means the high security of the S-box against differential approximation attacks. In Tables 11 and 12, we can see that our randomly picked S-box-1 and S-box-2 fully fill the DP criterion.

Table 11: DP of S-box-1.

| 0.00000 | 0.02343 | 0.03125 | 0.03125 | 0.02343 | 0.02343 | 0.02343 | 0.02343 | 0.03125 | 0.02343 | 0.02343 | 0.02343 | 0.02343 | 0.02343 | 0.02343 | 0.02343 | 0.03125 |
|---------|---------|---------|---------|---------|---------|---------|---------|---------|---------|---------|---------|---------|---------|---------|---------|---------|
| 0.03125 | 0.03125 | 0.03125 | 0.02343 | 0.02343 | 0.02343 | 0.03125 | 0.02343 | 0.02343 | 0.03125 | 0.02343 | 0.02343 | 0.03125 | 0.02343 | 0.02343 | 0.02343 | 0.03125 |
| 0.03125 | 0.03125 | 0.02343 | 0.03125 | 0.03125 | 0.03125 | 0.039062 | 0.02343 | 0.02343 | 0.02343 | 0.02343 | 0.02343 | 0.02343 | 0.02343 | 0.02343 | 0.02343 | 0.03125 |
| 0.02343 | 0.03125 | 0.02343 | 0.03125 | 0.03125 | 0.03125 | 0.02343 | 0.02343 | 0.02343 | 0.02343 | 0.02343 | 0.02343 | 0.02343 | 0.03125 | 0.02343 | 0.02343 | 0.03906 |
| 0.02343 | 0.02343 | 0.03125 | 0.02343 | 0.03125 | 0.02343 | 0.02343 | 0.02343 | 0.02343 | 0.03125 | 0.03125 | 0.03125 | 0.03125 | 0.02343 | 0.03125 | 0.03125 | 0.02343 |
| 0.03125 | 0.03125 | 0.03125 | 0.02343 | 0.02343 | 0.02343 | 0.03125 | 0.02343 | 0.03125 | 0.03125 | 0.02343 | 0.03125 | 0.02343 | 0.03125 | 0.02343 | 0.03125 | 0.03125 |
| 0.02343 | 0.03125 | 0.02343 | 0.03125 | 0.01562 | 0.03125 | 0.02343 | 0.02343 | 0.02343 | 0.02343 | 0.02343 | 0.01562 | 0.02343 | 0.02343 | 0.01562 | 0.02343 | 0.02343 |
| 0.02343 | 0.02343 | 0.02343 | 0.02343 | 0.03125 | 0.03125 | 0.01562 | 0.02343 | 0.02343 | 0.02343 | 0.03906 | 0.03125 | 0.02343 | 0.03906 | 0.03125 | 0.02343 | 0.03125 |
| 0.03125 | 0.02343 | 0.02343 | 0.02343 | 0.03125 | 0.03125 | 0.02343 | 0.02343 | 0.02343 | 0.02343 | 0.02343 | 0.02343 | 0.02343 | 0.02343 | 0.02343 | 0.02343 | 0.02343 |
| 0.03125 | 0.02343 | 0.03125 | 0.03125 | 0.02343 | 0.02343 | 0.03125 | 0.02343 | 0.02343 | 0.02343 | 0.02343 | 0.02343 | 0.02343 | 0.02343 | 0.02343 | 0.02343 | 0.02343 |
| 0.02343 | 0.02343 | 0.02343 | 0.02343 | 0.02343 | 0.02343 | 0.02343 | 0.02343 | 0.02343 | 0.02343 | 0.02343 | 0.02343 | 0.02343 | 0.02343 | 0.02343 | 0.02343 | 0.02343 |
| 0.02343 | 0.03125 | 0.02343 | 0.02343 | 0.03125 | 0.02343 | 0.03125 | 0.02343 | 0.02343 | 0.02343 | 0.02343 | 0.02343 | 0.03125 | 0.02343 | 0.02343 | 0.02343 | 0.02343 |
| 0.03125 | 0.02343 | 0.02343 | 0.02343 | 0.02343 | 0.02343 | 0.02343 | 0.02343 | 0.02343 | 0.03125 | 0.02343 | 0.02343 | 0.03125 | 0.02343 | 0.03125 | 0.03125 | 0.03125 |
| 0.02343 | 0.02343 | 0.02343 | 0.03125 | 0.03125 | 0.03125 | 0.02343 | 0.03125 | 0.02343 | 0.03906 | 0.02343 | 0.02343 | 0.03125 | 0.02343 | 0.03906 | 0.03125 | 0.02343 |
| 0.01562 | 0.02343 | 0.03125 | 0.02343 | 0.02343 | 0.02343 | 0.02343 | 0.03125 | 0.02343 | 0.02343 | 0.02343 | 0.03906 | 0.02343 | 0.01562 | 0.02343 | 0.02343 | 0.02343 |
| 0.02343 | 0.02343 | 0.03125 | 0.02343 | 0.02343 | 0.02343 | 0.02343 | 0.03125 | 0.03125 | 0.02343 | 0.02343 | 0.02343 | 0.02343 | 0.02343 | 0.03906 | 0.03125 | 0.02343 |
| 0.02343 | 0.02343 | 0.02343 | 0.02343 | 0.02343 | 0.02343 | 0.02343 | 0.03125 | 0.03125 | 0.03125 | 0.02343 | 0.02343 | 0.03125 | 0.02343 | 0.02343 | 0.01562 | 0.02343 |

TABLE 12: DP of S-box-2.

| | | | | | | | | | | | | | | | |
|---|---|---|---|---|---|---|---|---|---|---|---|---|---|---|---|
| 0.00000 | 0.02343 | 0.02343 | 0.03125 | 0.02343 | 0.015625 | 0.02343 | 0.03125 | 0.03125 | 0.02343 | 0.03125 | 0.02343 | 0.02343 | 0.02343 | 0.02343 | 0.02343 |
| 0.02343 | 0.02343 | 0.02343 | 0.02343 | 0.03125 | 0.02343 | 0.02343 | 0.02343 | 0.02343 | 0.03125 | 0.03125 | 0.03125 | 0.015625 | 0.02343 | 0.02343 | 0.02343 |
| 0.02343 | 0.02343 | 0.02343 | 0.02343 | 0.02343 | 0.03125 | 0.03125 | 0.02343 | 0.03125 | 0.02343 | 0.02343 | 0.02343 | 0.02343 | 0.02343 | 0.02343 | 0.02343 |
| 0.02343 | 0.03125 | 0.02343 | 0.02343 | 0.02343 | 0.03125 | 0.02343 | 0.03125 | 0.03125 | 0.03125 | 0.03125 | 0.02343 | 0.02343 | 0.02343 | 0.02343 | 0.02343 |
| 0.03125 | 0.02343 | 0.03125 | 0.03906 | 0.03125 | 0.03125 | 0.02343 | 0.03125 | 0.03125 | 0.03906 | 0.03125 | 0.02343 | 0.02343 | 0.03125 | 0.02343 | 0.02343 |
| 0.03125 | 0.02343 | 0.02343 | 0.03125 | 0.02343 | 0.02343 | 0.03125 | 0.02343 | 0.02343 | 0.03125 | 0.03125 | 0.02343 | 0.02343 | 0.02343 | 0.03125 | 0.02343 |
| 0.02343 | 0.02343 | 0.02343 | 0.02343 | 0.03125 | 0.03125 | 0.03125 | 0.03125 | 0.02343 | 0.03125 | 0.02343 | 0.03125 | 0.02343 | 0.02343 | 0.03125 | 0.03125 |
| 0.02343 | 0.02343 | 0.03125 | 0.03125 | 0.02343 | 0.02343 | 0.02343 | 0.02343 | 0.02343 | 0.03125 | 0.02343 | 0.03125 | 0.02343 | 0.02343 | 0.02343 | 0.02343 |
| 0.02343 | 0.02343 | 0.02343 | 0.02343 | 0.03906 | 0.03906 | 0.02343 | 0.02343 | 0.02343 | 0.03125 | 0.03906 | 0.02343 | 0.02343 | 0.02343 | 0.02343 | 0.02343 |
| 0.02343 | 0.03125 | 0.02343 | 0.02343 | 0.02343 | 0.02343 | 0.02343 | 0.02343 | 0.03125 | 0.02343 | 0.03125 | 0.02343 | 0.02343 | 0.02343 | 0.02343 | 0.03125 |
| 0.02343 | 0.02343 | 0.02343 | 0.02343 | 0.02343 | 0.03125 | 0.02343 | 0.03125 | 0.03125 | 0.02343 | 0.03125 | 0.03125 | 0.02343 | 0.02343 | 0.02343 | 0.02343 |
| 0.02343 | 0.03125 | 0.02343 | 0.02343 | 0.02343 | 0.03125 | 0.02343 | 0.03125 | 0.03125 | 0.02343 | 0.03125 | 0.02343 | 0.02343 | 0.02343 | 0.02343 | 0.03125 |
| 0.02343 | 0.02343 | 0.02343 | 0.02343 | 0.02343 | 0.02343 | 0.02343 | 0.03125 | 0.03125 | 0.02343 | 0.03125 | 0.02343 | 0.02343 | 0.015625 | 0.02343 | 0.02343 |
| 0.02343 | 0.02343 | 0.02343 | 0.03125 | 0.02343 | 0.02343 | 0.02343 | 0.02343 | 0.03125 | 0.02343 | 0.03125 | 0.02343 | 0.02343 | 0.02343 | 0.02343 | 0.03125 |
| 0.03125 | 0.02343 | 0.03125 | 0.03125 | 0.02343 | 0.03125 | 0.03125 | 0.03125 | 0.02343 | 0.03906 | 0.02343 | 0.02343 | 0.02343 | 0.02343 | 0.015625 | 0.03125 |
| 0.02343 | 0.02343 | 0.02343 | 0.03906 | 0.02343 | 0.02343 | 0.03125 | 0.03125 | 0.02343 | 0.03125 | 0.03906 | 0.02343 | 0.03125 | 0.03125 | 0.02343 | 0.03906 |

## 6. Conclusion

The protection of confidential information is a global issue, and block encryption algorithms are the most reliable option. The actual strength of the block encryption algorithms against several attacks is entirely dependent on S-boxes. Currently in the literature, algebraic structure-based and chaos-based techniques are highly used for the construction of S-boxes because both these techniques have favourable features for S-box construction, but many attacks of these techniques have been identified. In this paper, we purposed a novel technique for the dynamic generation of S-boxes that is safe against the existing attacks of algebraic structure-based and chaos-based techniques. True randomness has been universally recognized as the ideal method for security primitive because true random numbers are unpredictable, irreversible, and unreproducible. Rather than algebraic structure- and chaos-based techniques, our proposed technique depends on inevitable high-quality randomness which exists in underwater acoustics waves. According to our knowledge first time, this type of research has been conducted, in which natural randomness of underwater acoustic waves and knight's tour problem has been used for the generation of block cipher's substitution box. The proposed method satisfies all standard evaluation tests of S-boxes construction and true random numbers generation. Additionally, our dynamically generated S-boxes have better or equal strength, over the latest published S-boxes (2020 to 2021). In the future, we will extend this research for automatic key generation and optimization using knight's tour.

## Data Availability

The datasets analyzed during the current study are available in the Australian Antarctic Data Centre repository at https://data.aad.gov.au/metadata/records/AAS_4102_longTermAcousticRecordings.

## Conflicts of Interest

The authors declare no conflicts of interest.

## Acknowledgments

The authors extend their gratitude to the Deanship of Scientific Research at King Khalid University for funding this work through research groups program under grant number R. G. P. 2/132/42.